\documentclass[aps,twocolumn, floats,preprintnumbers,showpacs,nofootinbib,prd]{revtex4-1}
\pdfoutput=1
\usepackage{hyperref}
\usepackage{graphicx}
\usepackage{url}
\usepackage{amsmath}
\usepackage{amsfonts}
\usepackage{amssymb}
\def\beq{\begin{equation}}
\def\eeq{\end{equation}}
\def\beqa{\begin{eqnarray}}
\def\eeqa{\end{eqnarray}}

\def\ltap{\ \raise.3ex\hbox{$<$\kern-.75em\lower1ex\hbox{$\sim$}}\ }
\def\gtap{\ \raise.3ex\hbox{$>$\kern-.75em\lower1ex\hbox{$\sim$}}\ }

\begin{document}
\rightline{\vbox{\halign{&#\hfil\cr&SLAC-PUB-16027\cr}}}

\title{Coloron-assisted Leptoquarks at the LHC}

\author{Yang Bai\,$^{a}$ and Joshua Berger\,$^{b}$
\\
\vspace{2mm}$^{a}$ \normalsize\emph{Department of Physics, University of Wisconsin, Madison, WI 53706, USA}  \vspace{1mm} \\
$^{b}$ \normalsize\emph{SLAC National Accelerator Laboratory, 2575 Sand Hill Road, Menlo Park, CA 94025, USA}
}
%\vspace{2mm}
%\date{\today}

\pacs{12.60.-i,14.80.Sv}

\begin{abstract}
Recent searches for a first-generation leptoquark by the CMS
collaboration have shown around 2.5$\sigma$ deviations from Standard
Model predictions in both the $eejj$ and $e\nu jj$ channels.
Furthermore, the $eejj$ invariant mass distribution has another
2.8$\sigma$ excess from the CMS right-handed $W$ plus heavy neutrino
search.  We point out that additional leptoquark production from a
heavy coloron decay can provide a good explanation for all three
excesses.  The coloron has a mass around 2.1 TeV and the leptoquark
mass can vary from 550 GeV to 650 GeV.  A key prediction of this model
is an edge in the total $m_T$ distribution of $e\nu jj$ events at around
2.1 TeV.
\end{abstract}
\maketitle
%\keywords{}

%%%%%%%%%%%%%%%%%%%%%%%%%%%%%%%%
\noindent
{\it{\textbf{Introduction.}}}
%%%%%%%%%%%%%%%%%%%%%%%%%%%%%%%
First-generation leptoquark searches~\cite{CMS-leptoquark}
and right-handed $W$ gauge boson plus a heavy
neutrino searches~\cite{CMS:wprimenew} from the CMS collaboration have
both shown interesting deviations from the Standard Model (SM)
predictions in recent analyses. In this paper, we explore these
excesses and develop a potential new physics model that can explain them.

In the first-generation leptoquark searches, the CMS has studied 19.6
fb$^{-1}$ integrated luminosity of data at the 8 TeV LHC.  A
first-generation leptoquark~\cite{Pati:1974yy,Hewett:1997ce}, $S_1$ (in the notation of
Ref.~\cite{Buchmuller:1986zs}), can have two different decay channels,
$S_1 \rightarrow e^+ \bar{u}$ and $S_1 \rightarrow \nu_e \bar{d}$.
After they are pair-produced at the LHC via QCD interactions, the
final states at colliders are $e e jj$, $e\nu j j$ and $\nu\nu
jj$. The former two channels have been searched for and are reported
to deviate from the SM at 2.4$\sigma$ and 2.6$\sigma$ respectively
after imposing kinematic cuts to optimize a 650~GeV
leptoquark~\cite{CMS-leptoquark}.  For the $ee jj$ channel, in
addition to basic pre-selection, additional cuts are imposed on  the
scalar sum of the $p_{\rm T}$ of the two electrons and the two leading
jets, $S_{\rm T}$, the invariant mass of the two electrons, $m_{\rm
  ee}$ and the minimum of electron-jet invariant mass of the two
leptoquark candidates after choosing the combination with the smaller
difference between the two electron-jet masses, $m^{\rm min}_{ej}$.
The cuts optimized for a 650~GeV leptoquark are $S_{\rm T} > 850$~GeV,
$m_{\rm ee} > 155$~GeV and $m^{\rm min}_{ej} > 360$~GeV, for which
there are 36 observed events with $20.49\pm2.14\pm2.45 ({\rm syst})$
expected background events, which amounts to a 2.4$\sigma$ deviation
from the SM prediction. In the $e\nu jj$ channel, the missing transverse energy in the event,
$E^{\rm miss}_{\rm T}$, and the electron-neutrino transverse mass,
$m_{\rm T, e\nu}$, are also used to select events.  After imposing the
cuts, $S_{\rm T} > 1040$~GeV, $E^{\rm miss}_{\rm T} > 145$~GeV,
$m_{\rm ej} > 555$~GeV and  $m_{\rm T, e\nu} > 270$~GeV, there are 18
observed events in contrast to of $7.54\pm 1.20\pm 1.07 ({\rm syst})$
expected background events, representing a $2.6\sigma$ excess over the SM
prediction.  

In the $W^\pm_R$ plus a heavy neutrino $N_e$ search with 19.7 fb$^{-1}$
integrated luminosity at the 8 TeV LHC, a similar final state $e e jj$
has been used to probe $pp \rightarrow W_R \rightarrow e N_e
\rightarrow e e jj$. The signal selection cuts differ from the
cuts in the previous leptoquark searches. The cuts (beyond
pre-section) include $m_{\rm ee} > 200$~GeV and $m_{\rm eejj} >
600$~GeV~\cite{CMS:wprimenew}.  The invariant mass distribution of
$m_{\rm eejj}$ shows an excess at around 2 TeV.  For the bin from 1.8
TeV to 2.2 TeV, around 14 events have been observed with approximately
4.0 expected background events. Keeping only the statistical
uncertainty, this amounts to $2.8 \sigma$ local excess from the SM
prediction~\cite{CMS:wprimenew}.

Since the intriguing excesses in the $eejj$ channel happen in both
leptoquark and $W_R + N_e$ searches, the immediate question is whether
both excesses can be explained by the same model.  The
$W_R + N_e$ model cannot produce significant $e\nu jj$ events, so we
restrict ourselves to models producing two leptoquarks from a
combination of QCD and resonant production channels. We will later
comment on models with an event topology similar to the $W_R + N_e$
model.

Before introducing a detailed model and fitting to the data, we
introduce a few order of magnitude estimates regarding the data.  For
the leptoquark search, we consider the QCD-produced leptoquark model
and use it to give a rough sense of the excess, though other signal models
generally only have comparable acceptances at the order of magnitude level.  The
NLO QCD production cross-section for a 650 GeV leptoquark is
$13.2$~fb~\cite{Kramer:2004df}.  From Table 4  and Table 5 of
Ref.~\cite{CMS-leptoquark}, the leptoquark model predicts 125.85 and 37.22
events in the $eejj$ and $e\nu jj$ channels after the final selection
cuts, implying $48.6\%$ and $28.8\%$ signal acceptances, respectively.
Within the leptoquark model, one therefore obtains $\sigma(p p \to e e
j j) \sim 1.6$~fb for the $eejj$ channel and $\sigma(p p \to e \nu j j) \sim 1.9$~fb.  The
acceptance in the $W_R+N_e$ search is roughly independent of the chain
leading to the $eejj$ final state and indicates a production
cross-section of $\sigma(p p \to \mbox{resonance} \rightarrow e e j j)
\sim 1~{\rm fb}$, though this can include a contribution from
non-resonant production.  The similarity of these cross-sections
points to a common
origin for all three excesses, as well as electroweak symmetry
relations between the electron and neutrino signatures.  We explore
both of these possibilities in greater detail in this paper.

The remainder of this paper is structured as follows.  We begin by
introducing a coloron model that can be consistent with all current
data.  We then fit this model to the current excesses.  Given the model
details, we make several predictions for follow up searches.  
We conclude by briefly discussing some alternatives and their
distinguishing features.

%%%%%%%%%%%%%%%%%%%%%%%%%%%%%%%%
\noindent
{\it{\textbf{Coloron-assisted Leptoquark Model.}}}
\vspace*{1mm}
%%%%%%%%%%%%%%%%%%%%%%%%%%%%%%%%
Noting the approximately equal excesses in the $eejj$ and $e \nu jj$
channels, we consider a scalar leptoquark with $(\bar{3}, 1)_{1/3}$
under the SM gauge group. Following the notation in
Ref.~\cite{Buchmuller:1986zs}, we have the interaction of  
\beqa
g^{i j}_{1L} \bar{q}^{c\, i}_L \,i \tau_2\, \ell^j_L \, S_1 \,.
\eeqa
For the flavor assumption $g^{i j}_{1L} \approx
g^i_{1L} \delta^{ij}$ with $g^1_{1L} > g^2_{1L}, g^3_{1L}$, the
$S_1$ mainly couples to the first-generation quarks and
leptons. Because the $SU(2)_W$ symmetry, the leptoquark could decay
into $ej$ and $\nu_e j$ with equal branching ratios. Other operators
like $\bar{u}^c_R e_R S_1$ may break this branching ratio relation.
%$\bar{u}^c_R d_R S_1^\dagger$ by assuming small couplings

One simple extension of the leptoquark model which includes resonant
production is to introduce a coloron, which is a massive color-octet
gauge boson~\cite{Hill:1991at,Hill:1993hs,Chivukula:1996yr,Simmons:1996fz}. For
a simple two-site model with $SU(3)_1 \times SU(3)_2 \rightarrow
SU(3)_c$ from a Higgs mechanism, we have the massless gluon $G_\mu =
\cos{\theta}\,G_{1\,\mu} + \sin{\theta}\,G_{2\,\mu}$ and the massive
coloron $G^\prime_\mu = - \sin{\theta}\,G_{1\,\mu} +
\cos{\theta}\,G_{2\,\mu}$. The two gauge couplings satisfy $h_1
\cos{\theta} = g_s$ and $h_2 \sin{\theta} = g_s$ as well as $h_1/h_2 =
\tan{\theta}$. In this paper, we will ignore other potential
color-octet scalars in the renormalizable coloron model (see
Ref.~\cite{Bai:2010dj,Chivukula:2013xka,Chivukula:2014rka} for recent
studies). All the SM quarks couple to site number one, so one has the
coupling of $G^\prime$ to quarks
\beqa
g_s \, \tan{\theta}\,\bar{q} \,\gamma^\mu\, T^a\, G^{\prime a}_\mu \,q \,.
\eeqa
Depending on the site at which the leptoquark couples, one can have 
\beqa
i\,g_{S_1}\,g_s\,G^{\prime a}_\mu \left[ S_1 T^a \partial^\mu S_1^\dagger - (\partial^\mu S_1) T^{a} S_1^\dagger\right] \,,
\eeqa
with $g_{S_1} = \xi /\tan{\theta}$ for generalized to a multi-site model and
$S_1$ allowed to sit on a site beyond the two sites of $G^\prime$.
For $S_1$ just coupling to site number two, one has $\xi =
1.0$.  We will not consider the case with $S_1$ sitting on the site
number one as it cannot provide a sufficient signal cross section to
be an explanation for the observed excess. 

The coloron can decay into quarks as well as leptoquarks. The
partial decay widths of $G^\prime$ into the five light flavors, $t\bar t$
and leptoquarks are given by
\beqa
&&\Gamma(G^\prime \rightarrow jj) = \frac{5\,\alpha_s}{6}\,\tan^2{\theta} \,M_{G^\prime} \,, \\
&&\Gamma(G^\prime \rightarrow t\bar t) = \frac{\alpha_s}{6}\,\tan^2{\theta} \,M_{G^\prime} \left( 1 + \frac{2 m_t^2}{M^2_{G^\prime}} \right) \left(1 - \frac{4m_t^2}{M^2_{G^\prime}} \right)^{1/2}\,, \\
&&\Gamma(G^\prime \rightarrow S_1 S_1^\dagger) = \frac{g^{2}_{S_1}\,\alpha_s}{24} M_{G^\prime} \left( 1-\frac{4 M_{S_1}^2}{M^2_{G^\prime}} \right)^{3/2}  \,.
\eeqa
For the production of $G^\prime$, we can use the narrow width
approximation (for $0.15 < \tan{\theta} < 1/\sqrt{2}$,
$\Gamma_{G^\prime}/ M_{G^\prime} < 0.1$) to estimate the production
cross section for producing a $G^\prime$ in the $s$-channel:
\beqa
\sigma(q \bar{q} \rightarrow G^\prime) \approx \frac{8\pi^2\,\alpha_s\,\tan^2{\theta} }{9\,M_{G^\prime}} \delta\left(\sqrt{\hat s} - M_{G^\prime} \right) \,.
\eeqa
\begin{figure}[th!]
\begin{center}
\includegraphics[width=0.48\textwidth]{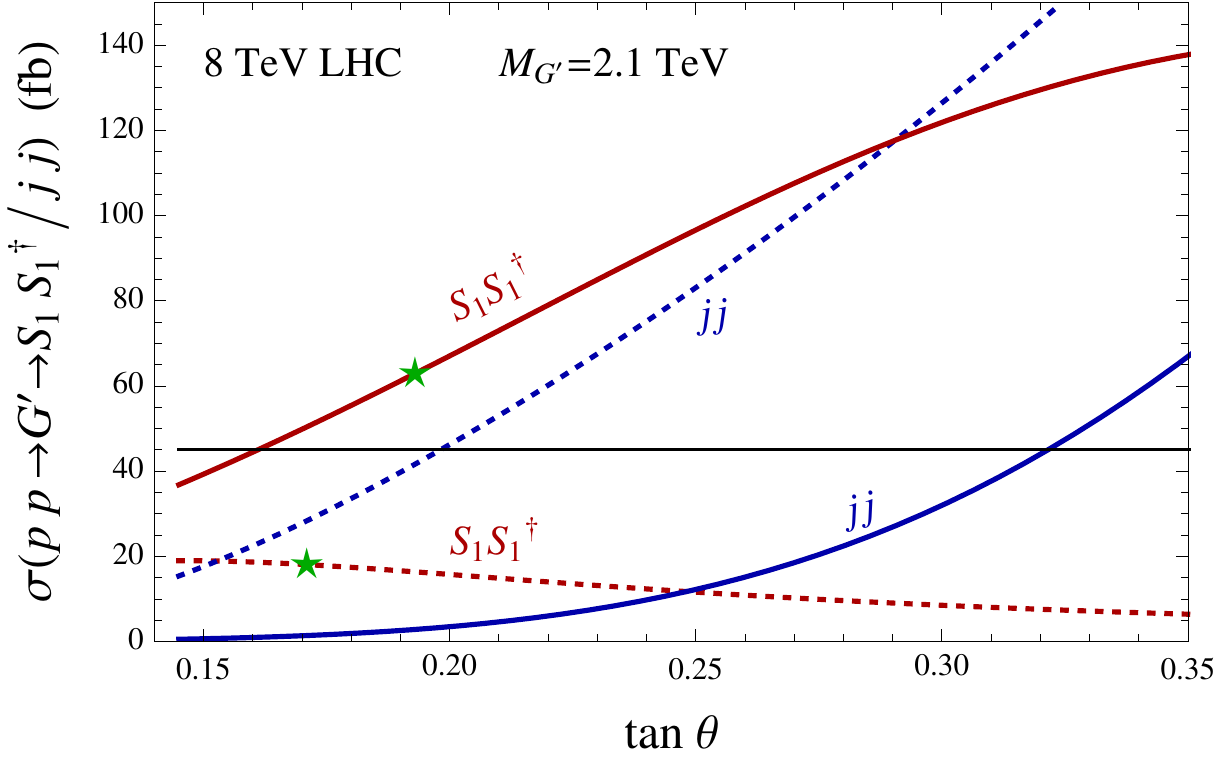}
\caption{The production cross sections of coloron times its various
  decay branching ratios. The solid lines have $M_{S_1}=550$~GeV with
  $\xi=1.0$, while the dotted lines have $M_{S_1}=650$~GeV with
  $\xi=0.15$. The black and horizontal line is the constraint from the
  narrow dijet resonance searches~\cite{CMS:dijet}. The two green
  five-pointed stars are the benchmark model points to fit the data.}
\label{fig:product}
\end{center}
\end{figure}

At the 8 TeV LHC and for $M_{G^\prime} = 2.1$~TeV (the location of the
most significant excess in the $eejj$ invariant mass
distribution~\cite{CMS:wprimenew}), the production 
cross section is $\sigma(pp \rightarrow G^\prime) \approx 1780 \times
\tan^2{\theta}$~fb. Using the MSTW~\cite{Martin:2009iq} PDFs as well
as the calculated branching ratios, we show $S_1 S_1^\dagger$ and $jj$
production cross sections from $G^\prime$ in
Fig.~\ref{fig:product}.  In the same plot, we also show the current 
constraints from dijet narrow resonance searches from 
CMS with 19.6 fb$^{-1}$ data. For the model with $\xi=1.0$ and
$M_{S_1}=550$~GeV the dijet has a constraint of $\tan{\theta} < 0.32$,
while for the model with $\xi=0.15$ and $M_{S_1}=650$~GeV the dijet
has a constraint of $\tan{\theta} < 0.19$ (see also
Ref.~\cite{Dobrescu:2013cmh} for more constraints on
other coloron masses without a leptoquark). The current $t\bar t$ resonance
searches~\cite{ATLAS-ttbar} are not sensitive enough to constrain the
model parameters in Fig.~\ref{fig:product}.

%--------------------------------------------------------------------------------
\noindent
{\it{\textbf{Fit to Data.}}}
%\vspace*{1mm}
%--------------------------------------------------------------------------------
We parametrize the model first with three phenomenological parameters,
$\sigma_{\rm SG} \equiv \sigma(pp \rightarrow G^\prime \rightarrow S_1
S_1^\dagger)$, $\mbox{Br}_{\rm ej}\equiv \mbox{Br}(S_1 \rightarrow e
j)$ and $\mbox{Br}_{\rm \nu j}\equiv \mbox{Br}(S_1 \rightarrow \nu j)$ to fit the three
excesses.  The signal acceptances for cases not studied in
\cite{CMS-leptoquark} are estimated by implementing the coloron model
in \verb+FeynRules+ \cite{Alloul:2013bka}, generating events at LO using
\verb+MadGraph+ \cite{Alwall:2014hca}, showering and hadronizing using
\verb+Pythia+ \cite{Sjostrand:2000wi}, and simulating the detector
using \verb+PGS+\cite{PGS}.  The selection cuts as outlined in \cite{CMS-leptoquark}
and \cite{CMS:wprimenew} are applied to the \verb+PGS+ events and the
signal acceptance is extracted.
\begin{table}[!tb]
  \renewcommand{\arraystretch}{1.6}
  \centering
  \begin{tabular}{c |  c |c |c |c}
    \hline     \hline
    LQ Mass& Production & LQ $eejj$ & LQ $\nu ejj$ & $W_R+N_e$ \\
    \hline
    550 (GeV)& QCD & 0.45 & 0.08 & 0.04 \\
     ~ & Coloron (2.1 TeV) & 0.60 & 0.18 & 0.55 \\
    \hline
    650 (GeV) & QCD & 0.49 & 0.29 & 0.08 \\
    ~ & Coloron (2.1 TeV) & 0.64 & 0.45 & 0.58 \\
    \hline     \hline
  \end{tabular}
  \caption{The acceptances for two benchmark leptoquark masses, for the
    three different searches, and for QCD and coloron-mediated productions.
    For the leptoquark searches, the
    acceptances are for the final selection of the cuts optimized for
    a $650$~GeV leptoquark~\cite{CMS-leptoquark}.  For the $W_R+N_e$ search, the acceptance is
    for the selected events to have $1.8~{\rm TeV} < m_{eejj} <
    2.2~{\rm TeV}$~\cite{CMS:wprimenew}.}
  \label{tab:acceptances}
\end{table}
The acceptances for two benchmark
leptoquark masses are shown in Table \ref{tab:acceptances}.  The
acceptances are for the final selection cuts optimized for a 650~GeV
leptoquark in the leptoquark searches and for selected events falling in the $1.8~{\rm TeV} <
m_{eejj} < 2.2~{\rm TeV}$ bin in the $W_R+N_e$ search.  Since
there are three searches and three parameters in 
this procedure, we solve for optimal parameters that fit the central
values of the excesses under the acceptances we calculated.  Taking
the coloron mass to be fixed at $2.1$~TeV, we find parameters
\beqa
\sigma_{\rm SG} = 63.0~\mbox{fb}\,,\quad \mbox{Br}_{\rm ej} = 0.12\,,\quad \mbox{Br}_{\rm \nu j} = 0.15 \,. 
\label{eq:fit-result-550}
\eeqa
for a leptoquark mass of 550~GeV and
\beqa
\sigma_{\rm SG} = 17.8~\mbox{fb}\,,\quad \mbox{Br}_{\rm ej} = 0.21\,,\quad \mbox{Br}_{\rm \nu j} = 0.13 \,. 
\label{eq:fit-result-650}
\eeqa
for a leptoquark mass of 650~GeV.  A $\chi^2$ fit shows that the model
with leptoquark mass 550~GeV is consistent with $\mbox{Br}_{\rm ej} = \mbox{Br}_{\rm
  \nu j}$, while the model with leptoquark 650~GeV is consistent with
$\mbox{Br}_{\rm ej} = 2\,\mbox{Br}_{\rm \nu j}$.  Either scenario is a
plausible result of electroweak symmetry. In terms of the parameter
$\tan{\theta}$ and from Fig.~\ref{fig:product}, the required
production cross sections can match to $\tan{\theta}=0.19$ and
$\tan{\theta}=0.17$ for $M_{S_1}=550$~GeV and $M_{S_1}=650$~GeV,
respectively.

\begin{figure}[th!]
\begin{center}
\includegraphics[width=0.44\textwidth]{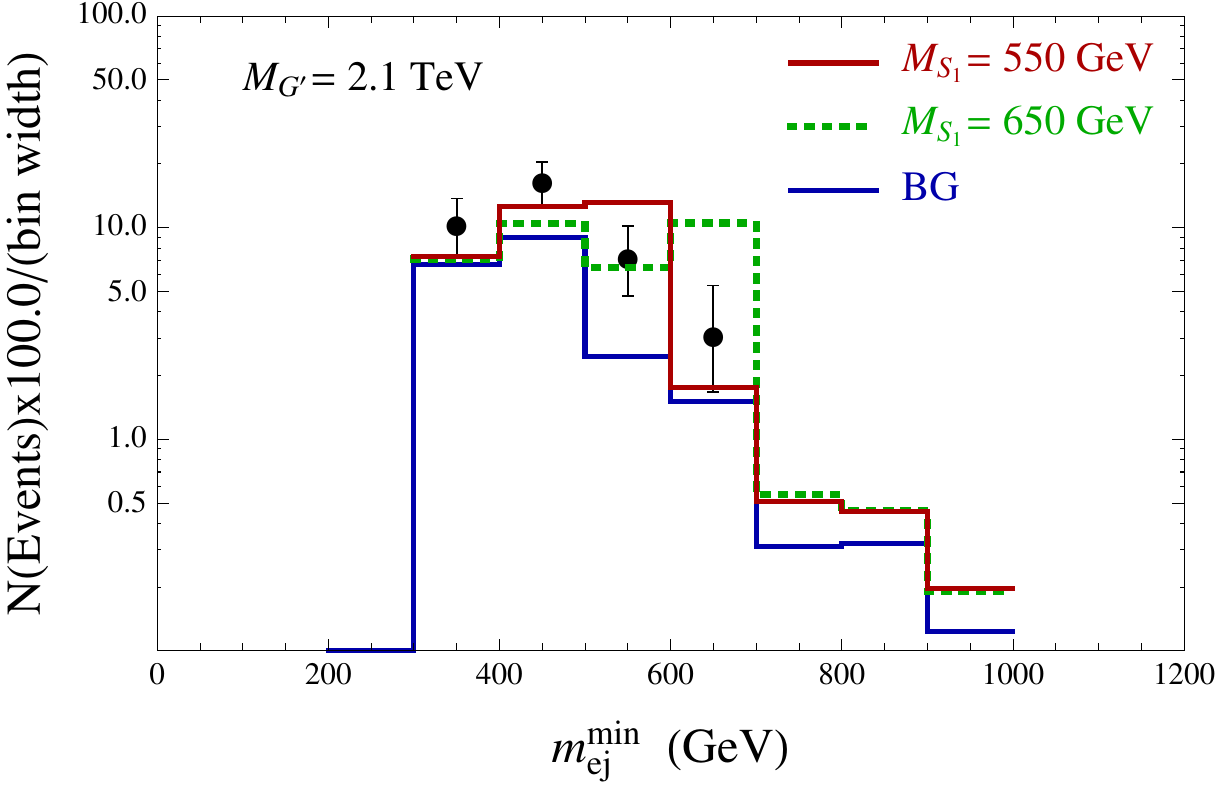}
\caption{A comparison of the data and the signal plus background $m_{\rm ej}^{\rm min}$ distributions from the leptoquark search in the $eejj$ final state. The fitted results in Eq.~(\ref{eq:fit-result-550}) and Eq.~(\ref{eq:fit-result-650}) are used for two benchmark leptoquark masses. The data and the SM background are taken from~\cite{CMS-leptoquark}.}
\label{fig:mejmin650}
\end{center}
\end{figure}

\begin{figure}[th!]
\begin{center}
\includegraphics[width=0.44\textwidth]{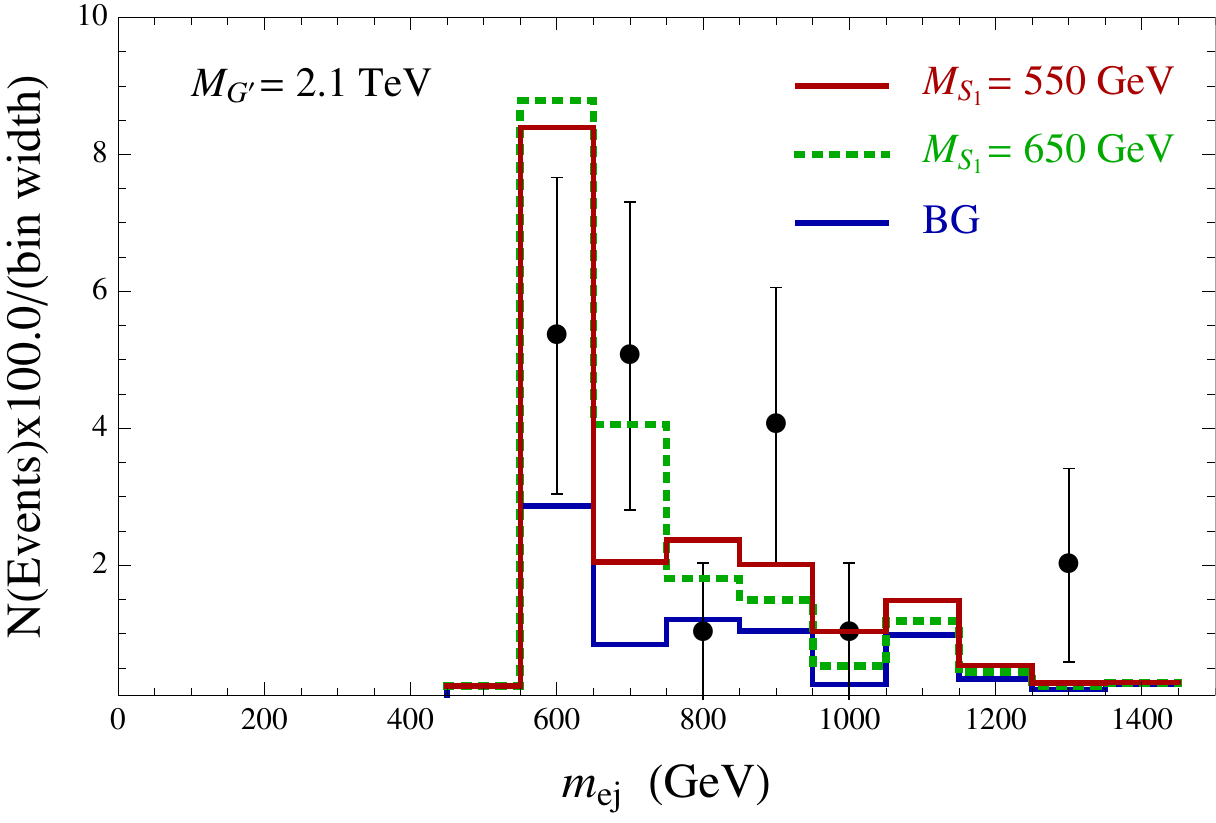}
\caption{The same as Fig.~\ref{fig:mejmin650} but in terms of $m_{\rm ej}$ in the $e\nu jj$ channel of the leptoquark search.}
\label{fig:mejnucut650}
\end{center}
\end{figure}

\begin{figure}[th!]
\begin{center}
\includegraphics[width=0.44\textwidth]{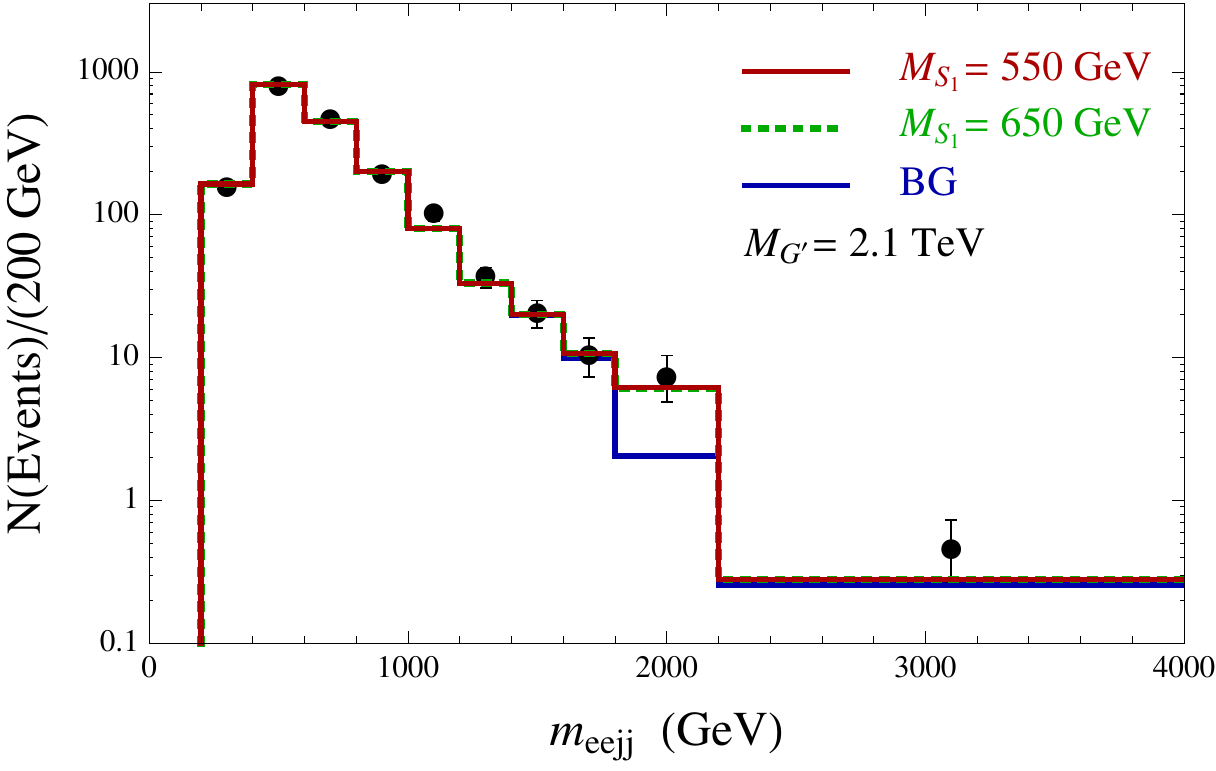}
\caption{The same as Fig.~\ref{fig:mejmin650} but in terms of $m_{\rm
    eejj}$ after imposing cuts used in the $W_R+N_e$ search. The data
  and the SM background are taken from~\cite{CMS:wprimenew}.}
\label{fig:meejjwrcut}
\end{center}
\end{figure}

Although we only use the total excess numbers of events to fit our model, we also show the $m_{ej}^{\rm min}$ distribution in the $eejj$ final state of the leptoquark search in Fig.~\ref{fig:mejmin650}, the $m_{ej}$ distribution in the $e\nu jj$ final state of the leptoquark search in Fig.~\ref{fig:mejnucut650} and the $m_{eejj}$ distribution in the $W_R+N_e$ search in Fig.~\ref{fig:meejjwrcut}. Comparing fitted results with two different leptoquark masses, one can see that the current data does not have enough statistics to constraint the leptoquark mass.

%

%--------------------------------------------------------------------------------
\noindent
{\it{\textbf{Predictions and Further Searches.}}}
%\vspace*{1mm}
%--------------------------------------------------------------------------------
These results have several implications for further searches, which we
now briefly outline.  Most obviously, the ATLAS experiment should be sensitive to any excesses in all three
channels studied here.  In addition, assuming the best fit coloron model,
ATLAS and CMS should see the following signatures:
\begin{itemize}
\item A bump in the $e \nu jj$ invariant mass distribution. Assuming a
  leptoquark mass, one can reconstruct events in this channel. If one
  cannot determine the leptoquark mass, one should still see an edge
  at $\sim 2.1$~TeV in the $m_T$ (constructed by summing the three visible
  particle transverse momenta and missing transverse momentum) distribution;
%In addition, the mass information known from the $ej$ spectrum could be used to solve for the neutrino momentum (up to a two fold degeneracy), possibly enhancing the signal;

\item A dijet+MET ($\nu\nu jj$) cross-section of $\sim 0.5 - 2.4$~fb
  depending on the scenarios.  The current limit on this signature is
  $17$~fb and $32$~fb in the two scenarios
  respectively~\cite{CMS-PAS-SUS-13-019}, 
  assuming the coloron production channel has the same acceptance as
  a squark.  This assumption is likely badly violated due to the
  harder objects in coloron events;

\item A dijet (or $t \bar t$) resonance with a mass $2.1$~TeV with $\sigma \times {\rm
    Br} \sim 1 - 20$~fb, again depending on the leptoquark mass scenarios.
\end{itemize}

The current searches can be improved to confirm our coloron+leptoquark
model. For example, one can find a bump in the invariant mass
distribution of $e+j$ pairs selected from events in the $2.1$~TeV peak
of the $W_R+N_e$ search. Additionally, events coming from a resonance
typically have a larger $S_T$, so a tighter $S_T$ cut would enhance the
signature of any resonant production model in any of the channels.

The above predictions are a required consequence of any incarnation of
the coloron model.  There are, however, other possible signatures that are more
dependent on the detailed structure of the model. Most importantly,
there must be other decay modes for the leptoquark, as the listed
branching fractions in Eqs.~(\ref{eq:fit-result-550}) and
(\ref{eq:fit-result-650}) to first generation leptons do not add up to
100\%.  Depending on the flavor model, the leptoquark can also decay
into other generations of leptons and quarks, for instance $S_1
\rightarrow \tau^+ \bar{t}, \nu_\tau \bar{b}$, which currently has
less stringent limits~\cite{Chatrchyan:2012sv,ATLAS:2013oea}. Simply
due to the leptoquark quantum numbers, other possibilities are
restricted.  The simplest ones include baryon-number 
violating couplings or additional $j + \mbox{MET}$ channels with the MET
from a pair of dark matter particles.  The former are constrained by the absence of
proton decay, while the later has no stringent constraints so far and will be probed by the dijet+MET search.
%the latter are constrained by the instability of
%the missing particle if its mass is larger than 100s of MeV and by the
%$\nu ejj$ final state if its mass is small.  If the DM mass is large
%(and the couplings correspondingly small), then a second bump could
%appear in the $m_{T,\nu j}$ distribution of $\nu ejj$-like events.
More exotic channels are also possible, including cascades to
additional jets, but all final states will include jets along with possible
leptons and/or MET.

One final possibility hinted at by the data is that the leptoquark decay branching ratios
to electrons and neutrinos may be the same, indicating a coupling only
to the left-handed leptons.  To fully assess this possibility, one requires a more precise determination
of the masses.

%--------------------------------------------------------------------------------
\noindent
{\it{\textbf{Discussion and Conclusions.}}}
%\vspace*{1mm}
%--------------------------------------------------------------------------------
The coloron plus leptoquark model is one well-motivated possible
explanation for the observed excess, but other models may also fit the data and have
qualitatively different additional signatures.  For example, a model with the
decay topology of the $W_R+N_e$ model studied in \cite{CMS:wprimenew} can
capture the quantitative features of the $eejj$ data presented in
\cite{CMS-leptoquark} at the level of current uncertainties. In fact,
our simulated results of the $W_R+N_e$ model show a broad peak
structure in $m_{\rm ej}^{\rm min}$ after the selection cuts of the
leptoquark search~\cite{CMS-leptoquark}. There could exist other event
topologies to provide the similar signatures (see
Ref.~\cite{Abdullah:2014oaa} for more event topologies). A similar
model that added a $e\nu jj$ decay mode could account for the data in that
channel as well.  The construction of a specific model with this
topology is beyond the scope of this work. Nevertheless, this
quasi-degeneracy should be probed further by examining the various
possible resonant combinations of the final state particles
($\ell\ell$, $jj$, $\ell jj$ and $\ell\ell j$, as well as the leptoquark combination).

\vspace{3mm}
%%%%%%%%%%%%%%%%%%%%%%%%%%%%%%%%
%\section{Acknowledgements} 
%\label{sec:production}
{\it{\textbf{Acknowledgements.}}}
%\vspace*{1mm}
%%%%%%%%%%%%%%%%%%%%%%%%%%%%%%%
We thank John Chou, Bryan Dahmes, Jared Evans, Paddy Fox, Francesco Romeo, Francesco Santanastasio and Haijun Yang for useful discussion. YB is supported by the U. S. Department of Energy under the contract DE-FG-02-95ER40896.  SLAC is operated by Stanford University for the US Department of Energy under contract DE-AC02-76SF00515. We thank the Aspen Center for Physics, under NSF Grant No. PHY-1066293, where this work is generated and finished.

%%%%%%%%%%%%%%%%%%%%%%%%%%%%%%%%%%%
\providecommand{\href}[2]{#2}\begingroup\raggedright\endgroup
%%%%%%%%%%%%%%%%%%%%%%%%%%%%%%%%%%%


\begin{thebibliography}{10}

\bibitem{CMS-leptoquark}
{\bf CMS} Collaboration, {\it {Search for Pair-production of First Generation
  Scalar Leptoquarks in pp Collisions at sqrt s = 8 TeV}},  Tech. Rep.
  CMS-PAS-EXO-12-041, CERN, Geneva, 2014.

\bibitem{CMS:wprimenew}
{\bf CMS} Collaboration, V.~Khachatryan {\em et.~al.}, {\it {Search for heavy
  neutrinos and W bosons with right-handed couplings in proton-proton
  collisions at sqrt(s) = 8 TeV}},
  \href{http://xxx.lanl.gov/abs/1407.3683}{{\tt arXiv:1407.3683}}.

\bibitem{Pati:1974yy}
J.~C. Pati and A.~Salam, {\it {Lepton Number as the Fourth Color}},  {\em
  Phys.Rev.} {\bf D10} (1974) 275--289.

\bibitem{Hewett:1997ce}
J.~L. Hewett and T.~G. Rizzo, {\it {Much ado about leptoquarks: A Comprehensive
  analysis}},  {\em Phys.Rev.} {\bf D56} (1997) 5709--5724,
  [\href{http://xxx.lanl.gov/abs/hep-ph/9703337}{{\tt hep-ph/9703337}}].

\bibitem{Buchmuller:1986zs}
W.~Buchmuller, R.~Ruckl, and D.~Wyler, {\it {Leptoquarks in Lepton - Quark
  Collisions}},  {\em Phys.Lett.} {\bf B191} (1987) 442--448.

\bibitem{Kramer:2004df}
M.~Kramer, T.~Plehn, M.~Spira, and P.~Zerwas, {\it {Pair production of scalar
  leptoquarks at the CERN LHC}},  {\em Phys.Rev.} {\bf D71} (2005) 057503,
  [\href{http://xxx.lanl.gov/abs/hep-ph/0411038}{{\tt hep-ph/0411038}}].

\bibitem{Hill:1991at}
C.~T. Hill, {\it {Topcolor: Top quark condensation in a gauge extension of the
  standard model}},  {\em Phys.Lett.} {\bf B266} (1991) 419--424.

\bibitem{Hill:1993hs}
C.~T. Hill and S.~J. Parke, {\it {Top production: Sensitivity to new physics}},
   {\em Phys.Rev.} {\bf D49} (1994) 4454--4462,
  [\href{http://xxx.lanl.gov/abs/hep-ph/9312324}{{\tt hep-ph/9312324}}].

\bibitem{Chivukula:1996yr}
R.~Chivukula, A.~G. Cohen, and E.~H. Simmons, {\it {New strong interactions at
  the Tevatron?}},  {\em Phys.Lett.} {\bf B380} (1996) 92--98,
  [\href{http://xxx.lanl.gov/abs/hep-ph/9603311}{{\tt hep-ph/9603311}}].

\bibitem{Simmons:1996fz}
E.~H. Simmons, {\it {Coloron phenomenology}},  {\em Phys.Rev.} {\bf D55} (1997)
  1678--1683, [\href{http://xxx.lanl.gov/abs/hep-ph/9608269}{{\tt
  hep-ph/9608269}}].

\bibitem{Bai:2010dj}
Y.~Bai and B.~A. Dobrescu, {\it {Heavy octets and Tevatron signals with three
  or four b jets}},  {\em JHEP} {\bf 1107} (2011) 100,
  [\href{http://xxx.lanl.gov/abs/1012.5814}{{\tt arXiv:1012.5814}}].

\bibitem{Chivukula:2013xka}
R.~S. Chivukula, A.~Farzinnia, J.~Ren, and E.~H. Simmons, {\it {Constraints on
  the Scalar Sector of the Renormalizable Coloron Model}},  {\em Phys.Rev.}
  {\bf D88} (2013) 075020, [\href{http://xxx.lanl.gov/abs/1307.1064}{{\tt
  arXiv:1307.1064}}].

\bibitem{Chivukula:2014rka}
R.~S. Chivukula, E.~H. Simmons, A.~Farzinnia, and J.~Ren, {\it {LHC Constraints
  on a Higgs Partner from an Extended Color Sector}},
  \href{http://xxx.lanl.gov/abs/1404.6590}{{\tt arXiv:1404.6590}}.

\bibitem{CMS:dijet}
{\bf CMS} Collaboration, {\it {Search for Narrow Resonances using the Dijet
  Mass Spectrum with 19.6fb-1 of pp Collisions at sqrts=8 TeV}}, .

\bibitem{Martin:2009iq}
A.~Martin, W.~Stirling, R.~Thorne, and G.~Watt, {\it {Parton distributions for
  the LHC}},  {\em Eur.Phys.J.} {\bf C63} (2009) 189--285,
  [\href{http://xxx.lanl.gov/abs/0901.0002}{{\tt arXiv:0901.0002}}].

\bibitem{Dobrescu:2013cmh}
B.~A. Dobrescu and F.~Yu, {\it {Coupling-mass mapping of dijet peak searches}},
   {\em Phys.Rev.} {\bf D88} (2013), no.~3 035021,
  [\href{http://xxx.lanl.gov/abs/1306.2629}{{\tt arXiv:1306.2629}}].

\bibitem{ATLAS-ttbar}
{\bf ATLAS} Collaboration, G.~Aad {\em et.~al.}, {\it {Search for $t\bar t$
  resonances in the lepton plus jets final state with ATLAS using 4.7 fb$^{-1}$
  of $pp$ collisions at $\sqrt{s} = 7$ TeV}},  {\em Phys.Rev.} {\bf D88}
  (2013), no.~1 012004, [\href{http://xxx.lanl.gov/abs/1305.2756}{{\tt
  arXiv:1305.2756}}].

\bibitem{Alloul:2013bka}
A.~Alloul, N.~D. Christensen, C.~Degrande, C.~Duhr, and B.~Fuks, {\it
  {FeynRules 2.0 - A complete toolbox for tree-level phenomenology}},  {\em
  Comput.Phys.Commun.} {\bf 185} (2014) 2250--2300,
  [\href{http://xxx.lanl.gov/abs/1310.1921}{{\tt arXiv:1310.1921}}].

\bibitem{Alwall:2014hca}
J.~Alwall, R.~Frederix, S.~Frixione, V.~Hirschi, F.~Maltoni, {\em et.~al.},
  {\it {The automated computation of tree-level and next-to-leading order
  differential cross sections, and their matching to parton shower
  simulations}},  \href{http://xxx.lanl.gov/abs/1405.0301}{{\tt
  arXiv:1405.0301}}.

\bibitem{Sjostrand:2000wi}
T.~Sjostrand, P.~Eden, C.~Friberg, L.~Lonnblad, G.~Miu, {\em et.~al.}, {\it
  {High-energy physics event generation with PYTHIA 6.1}},  {\em
  Comput.Phys.Commun.} {\bf 135} (2001) 238--259,
  [\href{http://xxx.lanl.gov/abs/hep-ph/0010017}{{\tt hep-ph/0010017}}].

\bibitem{PGS}
J.~S. Conway, {\it {Pretty Good Simulation of high-energy collisions}},
  \href{http://xxx.lanl.gov/abs/090401 release}{{\tt 090401 release}}.

\bibitem{CMS-PAS-SUS-13-019}
{\bf CMS} Collaboration, {\it {Search for supersymmetry in hadronic final
  states using MT2 with the CMS detector at sqrt(s) = 8 TeV}},  Tech. Rep.
  CMS-PAS-SUS-13-019, CERN, Geneva, 2014.

\bibitem{Chatrchyan:2012sv}
{\bf CMS} Collaboration, S.~Chatrchyan {\em et.~al.}, {\it {Search for pair
  production of third-generation leptoquarks and top squarks in $pp$ collisions
  at $\sqrt{s}=7$ TeV}},  {\em Phys.Rev.Lett.} {\bf 110} (2013) 081801,
  [\href{http://xxx.lanl.gov/abs/1210.5629}{{\tt arXiv:1210.5629}}].

\bibitem{ATLAS:2013oea}
{\bf ATLAS} Collaboration, G.~Aad {\em et.~al.}, {\it {Search for third
  generation scalar leptoquarks in pp collisions at $\sqrt{s}$ = 7 TeV with the
  ATLAS detector}},  {\em JHEP} {\bf 1306} (2013) 033,
  [\href{http://xxx.lanl.gov/abs/1303.0526}{{\tt arXiv:1303.0526}}].

\bibitem{Abdullah:2014oaa}
M.~Abdullah, E.~Albin, A.~DiFranzo, M.~Frate, C.~Pitcher, {\em et.~al.}, {\it
  {Systematically Searching for New Resonances at the Energy Frontier using
  Topological Models}},  {\em Phys.Rev.} {\bf D89} (2014) 095002,
  [\href{http://xxx.lanl.gov/abs/1401.1462}{{\tt arXiv:1401.1462}}].

\end{thebibliography}
\end{document}